# Volumetric Metamaterials versus Impedance Surfaces in Scattering Applications

S. Kosulnikov,[1,a] D. Filonov,[2] A. Boag,[1] and P. Ginzburg[1,2]

[1]*School of Electrical Engineering, Tel Aviv University, Tel Aviv 69978, Israel*

[2]*Center for Photonics and 2D Materials, Moscow Institute of Physics and Technology, Dolgoprudnyi 141700 Russia*

[a] Corresponding author: s.y.kosulnikov@gmail.com

**Abstract**

Artificially created media allow employing material parameters as additional valuable degrees of freedom in tailoring electromagnetic scattering. In particular, metamaterials with either negative permeability or permittivity allow creating deeply subwavelength resonant structures with relatively high scattering cross-sections. However, the equivalence principle allows replacing volumetric structures with properly designed curved impedance surfaces, ensuring the same electromagnetic properties. Here, we examine this statement from a practical standpoint, considering two structures, having a dipolar electric resonance at the same frequency. The first realization is based on arrays of inductively loaded electric dipoles printed on stacked circuit boards (a volumetric metamaterial), while the second structure utilizes a 4-wire spiral on a spherical surface (surface impedance realization). An intermediate conclusion is that the surface implementation tends to outperform the volumetric counterparts in the scenario when a single resonance is involved. However, in the case where multiple resonances are overlapping and lossy materials are involved, volumetric realization can have an advantage. The discussed structures are of significant importance to the field of electrically small antennas, superdirective antennas, and superscatterers, which find use in wireless communications and radar applications, to name just a few.

**Introduction**

Controlling scattering from an object is among main objectives of applied electromagnetic theory and related applications. Various approaches have been developed and employed in antenna design, and in certain cases, they are applied to enhance scattering cross-sections and improve directivities [1].

Enhancement of scattering cross-sections has a broad range of applications in many wireless communication systems. Increasing an object's visibility allows performing a reliable remote detection in, e.g., radio frequency identification (RFID) applications, airborne, and marine beacons-based monitoring, to name just a few. While corner reflectors are typically used to improve the visibility at high frequencies (typically X-band (8-12 GHz) and higher), those devices become bulky in MHz realizations. Increasing visibility for VHF (very high frequency band) and UHF radars (ultra high frequency band of 30-3000 MHz), capable of identifying distant objects even over the horizon, by using miniature scatterers might be beneficial.

Apart from its applied significance, the subject of super scattering keeps attracting attention to pathways of braking commonly accepted limits in antenna theory. Electrically small antennas, satisfying a condition '$ka<1$' ($k$ is a free space wavenumber, and $a$ is the smallest radius of a sphere, enclosing the antenna) have inherently small bandwidth and directivities, bounded from above by well-known Chu-Harrington [2], Geyi [3] limits and few other adjustments of those two. Scattering cross-sections of subwavelength structures obey a fundamental single-channel limit of $(2l' + 1)\lambda^2/(2\pi)$, where $\lambda$ is the free space wavelength, and $l'$ is related to the orbital angular momentum of a multipolar resonance ($l' = 1$ is the electric or magnetic dipolar case). It is worth noting that a vast majority of scenarios, dipolar resonances govern scattering properties. In the case of lossless structures, the value of a scattering cross-section is bounded from above by $3\lambda^2/(2\pi)$ [4–7]. In order to break this limit, several resonances within a subwavelength structure can be spectrally overlapped to increase either the scattering peak or its bandwidth. Similar ideas can be employed to increase the directivities of electrically small antennas. The so-called Einstein's needle radiation (extremely high directivity) can be obtained if a sufficient number of multipoles interfere constructively [8]. While several theoretical approaches to demonstrate this effect have been proposed, experimental realization remains extremely challenging. Practical limitations of a multiple resonances overlap approach have been discussed in the context of superdirective antennas and are related to tolerances in fabrication and internal losses within a practical device. Higher-order multipoles elevate the energy stored in the antenna near-field and, as a result, possess much higher ohmic losses in any practical case. Hence, a demonstration of new approaches to superdirective antenna designs and superscatterers has both fundamental and practical outcomes. Metamaterials hold a promise to bring new approaches into the field, as they allow considering material degrees of freedom as additional parameters for electromagnetic design. Assessing the capabilities of metamaterials to improve electromagnetic characteristics in the field of superscattering is the objective here.

Quite a few metamaterial designs have been investigated since the first demonstration on electromagnetic invisibility. The concept of a 'Cloaking Device' was introduced in 2006 [9,10] and has already gained considerable attention due to the emerging demand for invisibility in the radio frequencies [11], [12], [13] and for visible light [14–17]. The direct outcome of those approaches, motivating future developments in the field, is the requirement for materials with anisotropic, spatially varying electric and magnetic susceptibilities, which cannot be found in naturally available materials. However, electromagnetic properties of artificially created periodic composites with subwavelength nontrivially structured unit cells can be described in the frame of homogenization theories (e.g., [18], [19], [20]) that suggest possibilities to achieve quite peculiar material parameters. For example, negative refractive index composites ($\varepsilon, \mu < 0$) were realized with arrays of spatially organized split-ring resonators and copper wires [21–23]. Another realization of efficient wideband directive antennas with a hyperbolic metamaterial [24] is performed with a matching array of non-overlapping resonant elements [25].

While interactions of bulk metamaterials, subject to a plane wave illumination, are relatively well understood, performances of metamaterial-based scatterers are less investigated. The main complexity to relate performances of experimentally achievable structures to their homogenized counterparts comes from the demand to account for nontrivially shaped granular boundaries and a relatively small number of unit cells within the scatterer's volume – all those aspects are solely related to practical realizations. Nevertheless, an artificial magnon resonance in a deeply subwavelength metamaterial-based resonator has been recently demonstrated [26]. A spherical scatterer was shown to have a strong magnetic dipolar resonance at a frequency where effective permeability approaches the value of -2 (in an analogy with a localized plasmon resonance in a small sphere, which is obtained for ε≈-2). However, the drawback of this approach is the relative complexity of the experimental realization. In particular, the granularity of unit cells, forming the structure, and moderate tolerance in their fabrication, can significantly lower the scattering cross-section, which can stay significantly below the possible theoretical upper limit of $3\lambda^2/(2\pi)$. To reduce the impact of fabrication-related losses on scattering performances, one can consider an equivalent electromagnetic problem by utilizing one of the fundamental electromagnetic theorems, i.e., the surface equivalence principle [27]. It suggests the ability to replace a volumetric scatterer with an impedance surface, enclosing the volume of the initial structure. The new interior can be taken to be either a perfect electric conductor (practically a metal shell) (Love's principle, e.g., [27]) or an empty space. The knowledge of electromagnetic fields in the original problem allows calculating the required surface impedance, which can also be realized experimentally under certain circumstances. In the case of a dipolar electric resonance, this realization of an equivalent impedance surface is well known and is based on the so-called spherical helix antenna. The comprehensive analysis of subwavelength helical structure was reported in detail, e.g., [28].

The main goal of this work is to compare the electromagnetic characteristics of the volumetric dipolar scatterer and its counterpart, realized with the help of the equivalence surface impedance principle. Though those structures should provide the same responses from the theoretical standpoint, practical realizations might play a role. In particular, losses within metal elements, surface roughness, and other fabrication-related factors can significantly favor one realization over another – underlining the practical limitations of volumetric metamaterials and surface impedances is the objective of this investigation (yet in the context of a single resonance without co-locating several multipoles). Electromagnetic performances of the following structures will be compared: a metamaterial-based volumetric scatterer made of inductively loaded dipoles (Fig. 1(a)), a homogenized counterpart of the metamaterial structure (Fig. 1(b)), and a surface impedance equivalent – spherical helix (Fig. 1(c)).

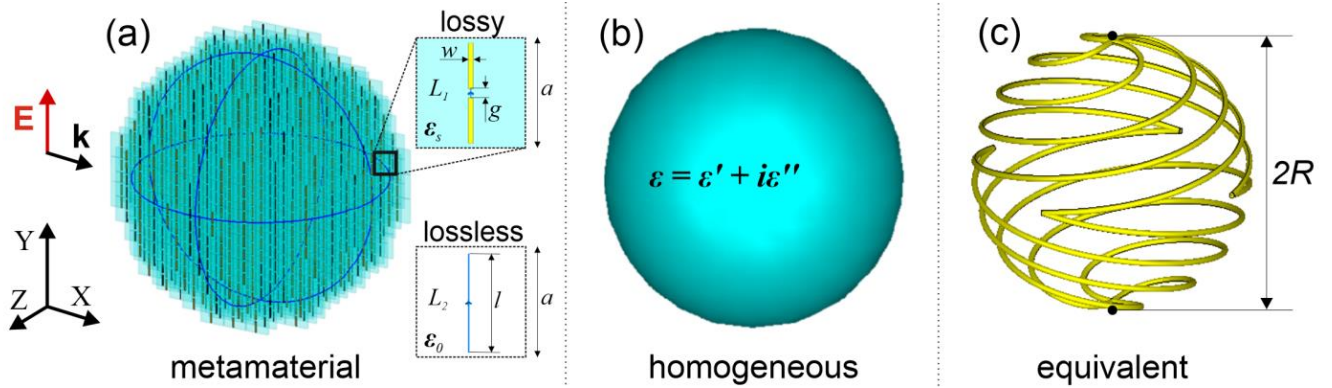

Fig.1. Electromagnetic scatterers. (a) Metamaterial-based volumetric scatterer made of inductively loaded dipoles. (b) Homogenized counterpart of the metamaterial structure. (c) Surface impedance equivalent – a 4-wire spherical helix.

Generally, the implementation of equivalent impedance on conformal surfaces is an extremely difficult practical task that does not have a unique solution. Our goal is to take the most simple and fundamental case – a spherical scatterer. This is a well-known and well-understood structure that has a complete closed-form analytical solution. Hence, it allows eliminating all unnecessary factors that might affect the comparison between curved impedance surfaces and volumetric metamaterials.

The manuscript is organized as follows: the metamaterial design is discussed first and then followed by an analysis of metamaterial-based scatterers. The surface equivalent (helical spiral) is analyzed next. A detailed comparison between electromagnetic responses of volumetric and surface structures comes before the Conclusion.

**Design of negative permittivity metamaterial**

Natural materials with negative permittivity at MHz-GHz spectral ranges are abundant in nature. Nevertheless, metamaterial counterparts can be constructed from ordered arrays of subwavelength resonators, e.g., [11]. Dipolar resonances of individual unit cells hybridize via near-field coupling and give rise to a collective mode associated with an effective susceptibility. If resonant responses are strong enough, effective permittivity ε of an array can cross zero and become negative. To achieve a strong resonance in a subwavelength dipole, the latter can be inductively loaded. In this case, the penalty comes in a bandwidth, which shrinks with the dipole's size decrease, according to the Chu-Harrington limit [29]. Further size reduction of a unit cell will imply having a requirement on high fabrication tolerance; otherwise, dipolar resonances in adjacent cells will not hybridize.

Our design is based on the methodology presented in [26]. The method assumes a two-port system where the ports are positioned along the direction of the propagating electromagnetic wave. The effective permittivity and permeability are extracted from the complex reflection and transmission coefficients. The structure of the unit cell appears in the inset to Fig. 1(a)). The operational frequency was chosen to be around 300 MHz, relevant to long-range radar applications. Furthermore, in this case, the unit cell sizes (should be much smaller than the operational wavelength, and quite a few ones should fit the scatterer to reduce the impact of granular boundaries) are appropriate for fabrication with standard printed board techniques and lumped elements' soldering. In the following studies, we will compare the performances of idealized lossless unit cells with their realistic counterparts, taking into account material losses within constitutive elements.

After a set of optimizations of the dipole sizes and the loading impedance, the following parameters of the unit cell were found: $l = 11$ mm is the entire length of the dipole; $w = 0.5$ mm is the width of the printed dipole strip; $g = 1$ mm is the dipole feeding gap; $t_{substr} = 1.5$ mm is the substrate thickness, $h_{metal} = 0.1$ mm is the thickness of the metal layer, $a = 12$ mm is the array periodicity in XZ-plane, and $a_y = 8$ mm is the period between the layers. The inductive load in the dipoles' gap is $L_1 = 2400$ nH for the lossy case and $L_2 = 8000$ nH for the lossless thin wire approximation case. The perspective view of the model is presented in Fig. 1(a). The thin blue lines show the contour of an imaginary 60 mm radius sphere, enclosing the structure. We used a metal with the electric conductivity $\sigma = 5.96 \cdot 10^7$ S/m, corresponding to "Copper (pure)" from the CST Studio Suite material library and Isola IS680 AG338 (with $\varepsilon = 3.38$ and tg $\delta = 0.0026$) for the substrate material to mimic volumetric losses of realistic metamaterials. In the case of lossless structure studies, a perfect electric conductor (PEC) as a material and a thin wire approximation for the dipole were used.

The results of the parametric retrieval, made with the waveguide configuration, appear in Fig. 2, demonstrating a range of frequencies, where negative permittivity is obtained. We used the Frequency Solver method with adaptive tetrahedral meshing in the CST Studio Suite as the most accurate solution for an electrically small geometry. An almost perfect complex Lorentzian shape of the effective medium dispersion can be observed for the realistic lossy case.

Lossless unit cells, however, possess much more complicated responses. Instead of a single hybridized mode, other high order contributions start playing a role. Higher order multipoles have stronger confinement and, hence, are usually suppressed by lossy materials. In the lossless case, however, they can contribute to extra-complexity of the homogenization procedure, e.g., spatial dispersion [20,30–32], which can also affect scattering performances, e.g., [17], [33]. Moreover, standard homogenization approaches (as the one used here) can fail to predict the behavior close to high-Q resonances. Nevertheless, we will utilize effective parameters in the lossless case by considering frequencies away from the resonance.

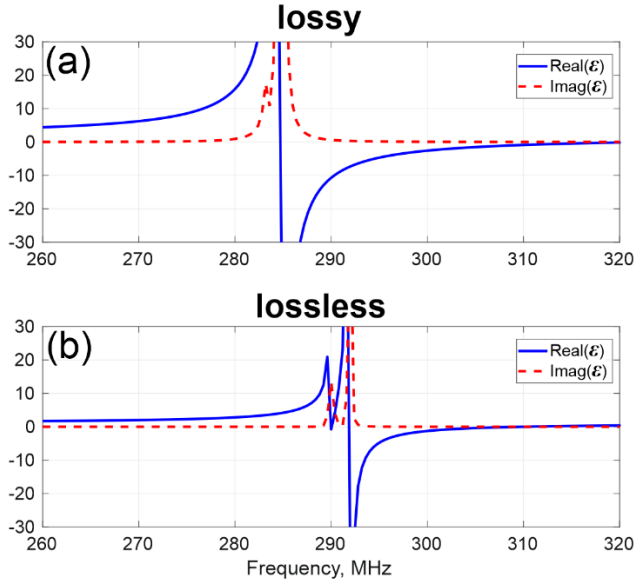

Fig. 2. Extracted material parameters dispersion. Effective permittivity component along the direction of the dipoles. (a) Lossy and (b) lossless (thin wire approximation) cases.

**Volumetric metamaterial scatterer**

The bulk metamaterial, investigated in the previous section, will be used to design scatterers with a finite extent. Here, in contrast with homogeneous materials, the granularity of the unit cell will affect scattering performances. In a theoretical case, where quite a few unit cells form the scatterer, the impact of modified boundary conditions on the scattering cross-section is not very high [34]. However, practical realizations must account for experimental aspects, suggesting reducing the number of unit cells within a scatterer and compromising on performances. The impact of the granularity (a small number of unit cells in a volume) on scattering cross-sections will be investigated next. We use the transient analysis with a varied step of hexahedral meshing in the CST Studio Suite software for these simulations.

Fig. 3 summarizes the results, comparing scattering cross-sections of metamaterial-based spheres with different radii with their homogenized counterparts (losses are fully accounted in both cases). The number of unit cells within a scatterer is proportional to its volume. It can be seen that only 3 unit cells fit the 10 mm radius structure. It is quite expected that scattering performances, in this case, are far from those predicted by Mie theory (red curves in Fig. 3). However, a 30 mm radius structure already comprises 125 unit cells. Its response reasonably fits the homogeneous case (note that scattering cross-section spectra are presented on a logarithmic scale; hence only orders of magnitude are visible for the comparison).

A further increase of the size and, as a result, elimination of the granularity impact on scattering allows approaching the fundamental single channel limit in scattering, represented with a dashed line on the figures. Metamaterial structure with $R = 60$ mm (will be used as a reference in the next section) is still below the single-channel limit; nevertheless, a relatively good spectral fit to Mie theory facilitates utilizing the homogenization models in this case. Fast oscillations are quite typical in numerical modeling, where the time-domain solver is applied to complex structures with moderately high-quality resonances. Electromagnetic energy remains within a structure, and residuals are bounces inside before they leave. Those fast oscillations can be cleaned by postprocessing, e.g., by applying a low pass filter. Here we did not do that, since the convergence is rather good.

It is worth noting that the metamaterial approach allows disentangling the size of the scatterer from metamaterial properties, if a sufficient number of unit cells fits inside the volume. It will be seen in the next section that the impedance surface approach might face difficulties in this type of scaling.

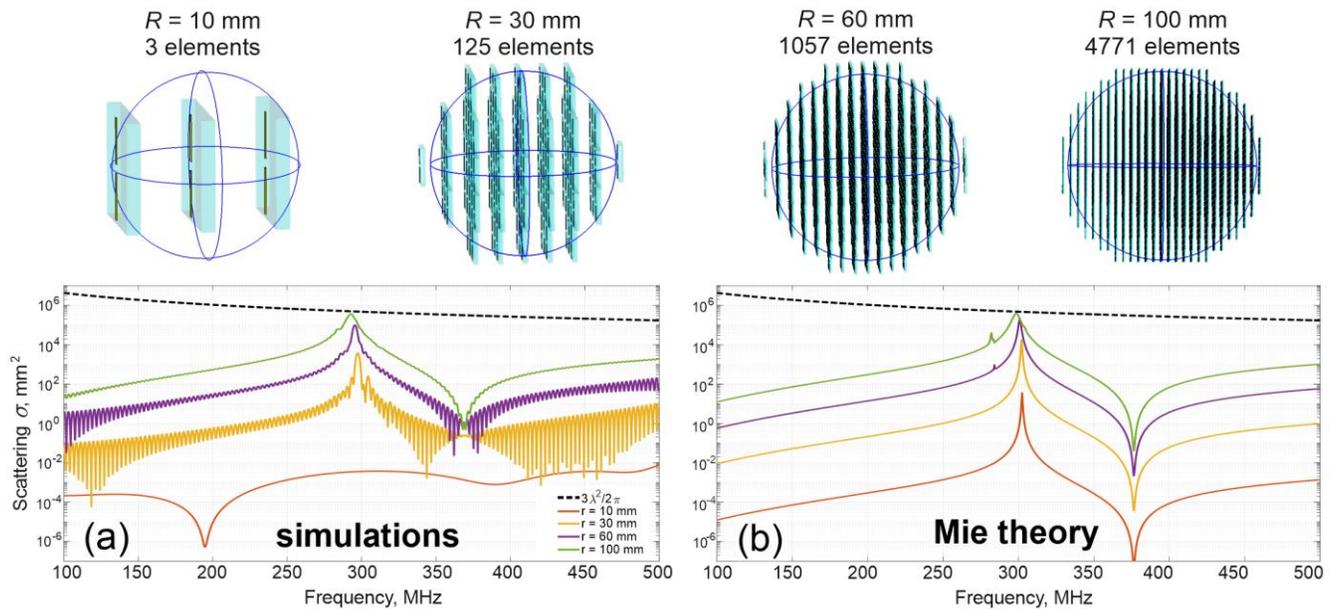

Fig. 3. Scattering cross-section spectra of subwavelength spheres, numerical analysis. (a) Metamaterial-based scatterers (geometries are in the upper insets). (b) Homogeneous spheres, with the parameters, retrieved from Fig. 2, lossy case). Different sizes correspond to solid color lines, elaborated in the captions. The dashed black line shows a theoretical single-channel limit for the dipolar scatterer. Insets are the granular metamaterial scatterers, analyzed in panel (a).

**Surface impedance-based scatterer**

Surface equivalent impedance can be found by applying discontinuity boundary conditions for electromagnetic fields, found by solving a full scattering problem. While those calculations provide quite generic results, practical implementation of the obtained impedances can be rather challenging. However, in relatively simple scenarios, surface equivalents can be realized in practice. For example, it has been shown that helical scatterers (Fig. 4(a)) operating at a fundamental dipolar resonance, are surface equivalents of subwavelength spheres with negative permittivity. Several designs of helixes were reported in [28,35], verifying this claim. The objective now is to compare those surface equivalents with metamaterials in terms of scattering performances, taking into account practical limitations.

Helical scatterers were constructed by using the formulation from [28,35]. Here, four identical wires were twisted around the structure's axis and mapped on an imaginary spherical surface (Fig. 4(a)). We used a Frequency Solver simulation method with a fairly dense tetrahedral meshing in the CST Studio Suite as a more accurate solution for this type of relatively not complex structure, which, however, requires to discretize the shape of the round cross-section of a helix-wire with the finite subwavelength radius. Available degrees of freedom for optimization are (i) the number of the wire's turns, (ii) the radius of the enclosing sphere, and (iii) the thickness of the wires, which has a minor impact on the electromagnetic response if the wires are kept to be relatively thin. It is worth noting that (i) and (ii) define the wires' overall length. To review the tuning capabilities, the number of turns will be kept constant (1 turn), whereas the radius of the enclosing sphere is varied. Table 1 summarizes the results:

Table 1. Resonance tuning of a spherical helix scatterer

| $R_{sphere}$, mm | $r_{wire}$, mm | Wire's length $L$, mm | Resonant frequency $f_0$, MHz | Resonant wavelength $\lambda$, mm | $2L/\lambda$ | $R/\lambda$ |
|---|---|---|---|---|---|---|
| 10 | 0,5 | 54,17 | 2011 | 149,18 | 0,73 | 0,067 |
| 30 | 1.32 | 162,5 | 663 | 452,49 | 0,72 | 0,067 |
| 60 | 1.32 | 324,99 | 315,8 | 949,97 | 0,68 | 0,063 |

It can be seen from Table 1 that the resonant conditions are mostly governed by the ratio $2L/\lambda$, which appears to be nearly constant. This kind of behavior is, indeed, expected from folded dipoles. Mutual inductance, emerging in folded geometries, introduces additional corrections to the resonant frequency, which is shifted in respect to $\lambda/2$ condition. For very small radii, the wires' thickness starts playing a role too.

Further tuning of the resonant frequency beyond the capabilities, summarized in Table 1, can be achieved by introducing additional capacitive and inductive loads. For example, introducing lumped elements into polar and equatorial points of the structure leads to the increment of the system's entire impedance. As a result, the resonant shifts to longer wavelengths. It is worth noting that the resonant tuning with a metamaterial allows moving the resonance to both higher and lower frequencies, which can be an advantage.

Figure 4 summarizes the results of scattering cross-section spectra for several geometries. The resonance tuning can be clearly seen along with the peaks, which reach a single-channel scattering limit, as it was predicted theoretically. Notably, material losses (copper has been used) almost do not affect the performances.

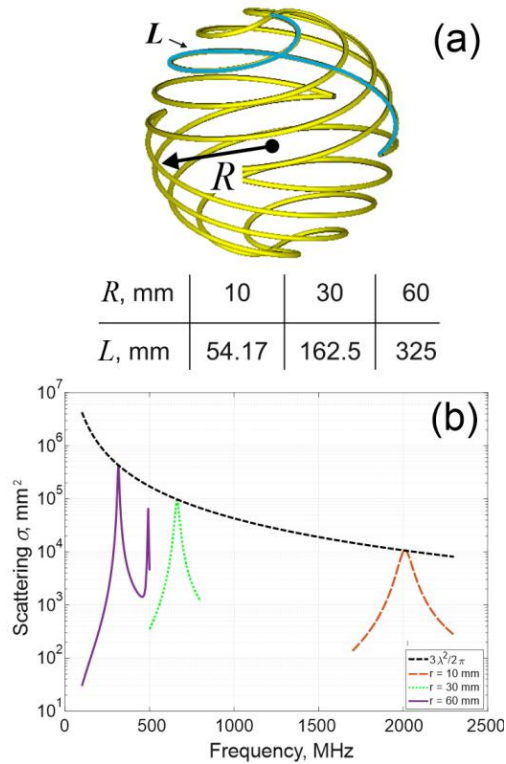

Fig. 4. (a) Schematics of the 4-wire helical scatterer and the parameters used in the numerical modeling. (b) Total scattering cross-section as the function of frequency for three different structures (the parameters are in panel (a), the wires' material is copper).

Practical implementation of helical scatterers can be based on a manual tilting of wires or with advanced additive manufacturing techniques, e.g., [36], [37].

**Comparison between metamaterials and impedance surfaces**

After discussing the electromagnetic performances of metamaterial-based scatterers and helical structures, their performances can be compared. From the theoretical standpoint, those structures are predicted to be equivalent, whereas practical aspects can set several restrictions that will be discussed hereinafter.

The considered structures were tuned to resonate at the same frequency band. Nevertheless, they possess certain deviations, one in respect to another. Table 2 summarizes the performances of seven different realizations. The '4-wire helix, lossless' is the helical scatterer made of PEC – as it will be seen later – it shows the best scattering performances. The '4-wire helix, lossy' is the same structure as the previous one, though the material is copper. The 'homogeneous sphere, lossless' is the sphere with material parameters corresponding to those in Fig. 2(a) with a nulled imaginary part of permittivity (those material parameters are not causal and taken for a reference only, The 'homogeneous sphere, lossy' is the same structure but with both real and imaginary permittivity given in Fig. 2(a). The 'metamaterial, lossless – the thin wire approximation' corresponds to free-standing PEC wires without a substrate, the 'metamaterial, lossless – dipoles on a lossless substrate' resemble practical realization in terms of the geometry, but the material losses are nulled, and the 'metamaterial, lossy' is the replica of the practical realization.

Table 2. Comparison of scattering performances between different spherical structures with all spheres' radius $R = 60$ mm.

| System | Parameters | Maximal scattering cross-section [m²] | Resonant frequency [MHz] |
|---|---|---|---|
| 4-wire helix, lossless | $r_{\text{wire}} = 1.32$ mm <br> $N_{\text{wire turns}} = 1$ <br> Material – PEC | 0.466 | 315.9 |
| 4-wire helix, lossy | $r_{\text{wire}} = 1.32$ mm <br> $N_{\text{wire turns}} = 1$ <br> Material - copper | 0.455 | 315.8 |
| Homogeneous sphere, lossless | $\varepsilon(f) = \varepsilon' + 0''$ <br> [extracted dispersion - Fig. 2(a)] | 0.505 | 300.9 |
| Homogeneous sphere, lossy | $\varepsilon(f) = \varepsilon' + \varepsilon''$ <br> [extracted dispersion - Fig. 2(a)] | 0.116 | 300.9 |
| Metamaterial, lossless – thin wire approximation | $l = 11$ mm <br> $w = 0$ mm | 0.520 | 283.9 |
| | $g = 1$ mm<br>$h_{metal} = 0.1$ mm<br>$a = 12$ mm<br>$a_y = 8$ mm<br>$L = 2400$ nH<br>PEC<br>No substrate | | |
| Metamaterial, lossless – dipoles on a lossless substrate | $l = 11$ mm<br>$w = 0.5$ mm<br>$g = 1$ mm<br>$h_{metal} = 0.1$ mm<br>$a = 12$ mm<br>$a_y = 8$ mm<br>$L = 2400$ nH<br>PEC<br>$t_{substr} = 1.5$ mm<br>Isola IS680 AG338 $\varepsilon' = 3.38$, tg($\delta$) = 0 | 0.157 | 295.6 |
| Metamaterial, lossy | $l = 11$ mm<br>$w = 0.5$ mm<br>$g = 1$ mm<br>$h_{metal} = 0.1$ mm<br>$a = 12$ mm<br>$a_y = 8$ mm<br>$L = 2400$ nH<br>Copper $\sigma = 5.96 \cdot 10^7$ S/m<br>$t_{substr} = 1.5$ mm<br>Isola IS680 AG338 $\varepsilon' = 3.38$, tg($\delta$) = 0.0026 | 0.1 | 295.6 |

In order to perform a fair comparison between the structures from Table 2, the following normalization per each scatterer has been applied: (i) the frequency axis was divided by a resonant frequency (i.e., each object resonates at $f_0$=1), (ii) the scattering cross-section is normalized to $\lambda_{res}^2$, where $\lambda_{res}$ is the resonant wavelength. The numerical results are summarized in Fig. 5 and compared to the theoretical single-channel scattering limit. The following observations can be done: spiral structures and idealized homogenous spherical scatterer reach the single-channel limit and slightly overcome it (it is a minor contribution of higher-order modes. However, metamaterial-based scatterers demonstrate several times weaker performances; nevertheless, they are at the same order of magnitude with the theoretical limit (recall Fig. 2, where the differences are almost unseen on the logarithmic scale). The homogenized lossy model of metamaterials predicts their behavior quite well. The results for the thin-wire representation of the volumetric structure can be considered in separate. Fast oscillations in the scattering-cross section spectrum are a manifestation of higher-order modes, which are not damped owing to vanishing losses. Similar behavior is observed in Fig, 2(b), where the homogenization procedure faced difficulties in retrieving effective material properties next to the resonance. Nevertheless, the scattering cross-section at the peak reaches the theoretical limit.

Apart from the scattering peak, the bandwidth also plays a critical role. Being subject to the Chu-Harrington limit, it is extracted from radiation reaction considerations, which completely neglect the dispersion. It can be seen in Fig. 5 that the bandwidth of all the structures is quite different; nevertheless, they all operate at the electrical resonance. While the surface equivalence principle predicts the same performances at a single frequency, aspects of dispersion are usually neglected. However, bandwidth consideration is important in many practical applications. In this context, metamaterials can provide an additional degree of freedom in designing electrically small and efficient scatterers. Notably, the homogeneous sphere simulations were performed with the Frequency Solver method and tetrahedral meshing in the CST Studio Suite (both for lossless and lossy cases) as it takes a more accurate result of the extracted material parameters than the Time Domain solver method (which takes parameters in accordance with its own approximation methods and might lead to an inaccurate solution). These results fully agree with the Mie theoretical analysis discussed in the previous section.

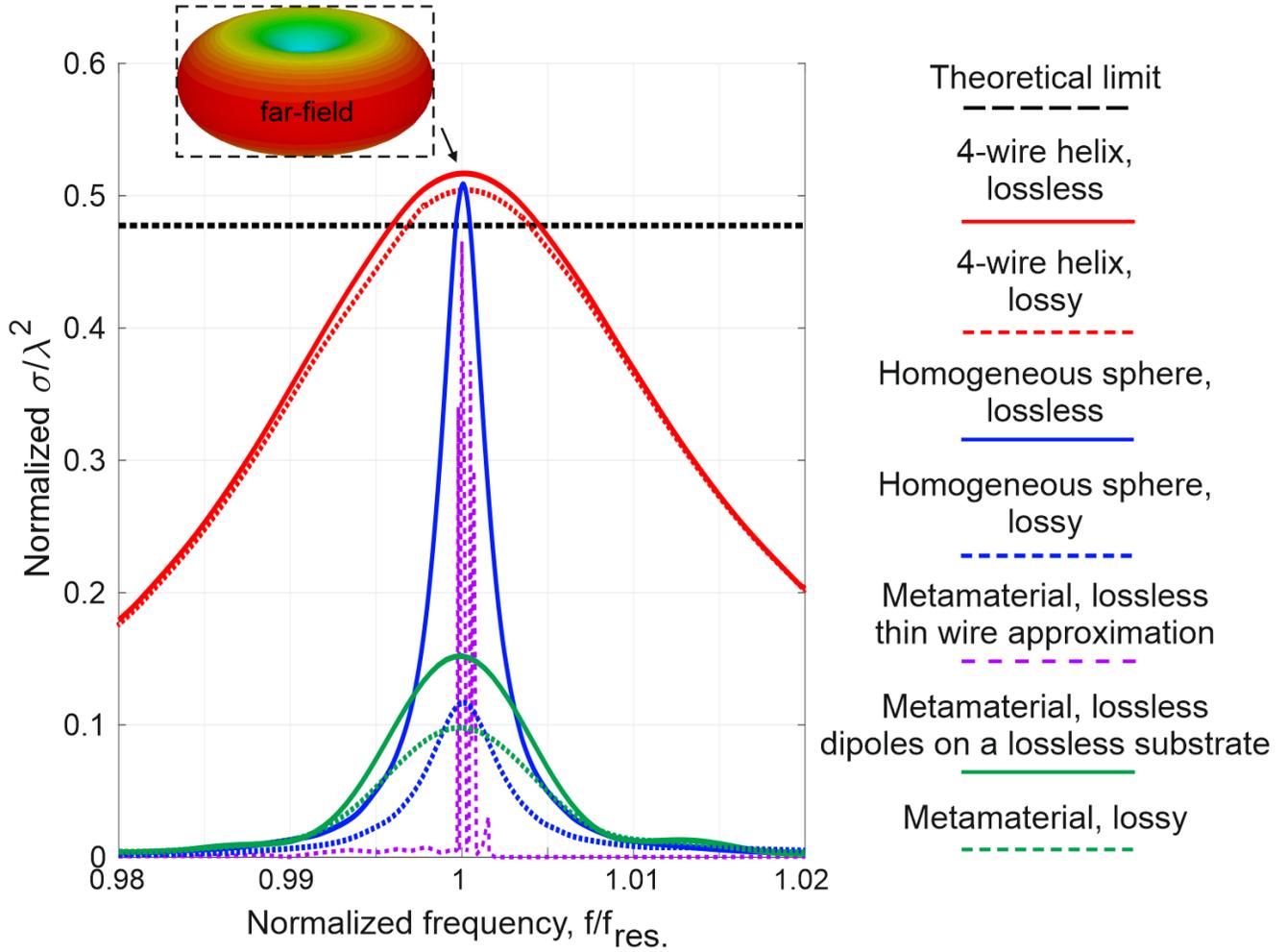

Fig.5. Comparison between the scattering performances of different objects indicated in the legend and Table 2. The frequency axis is normalized to a resonant frequency and the peak to a resonant wavelength squared per each scatterer individually. The dashed black line is the theoretical single-channel limit. The upper inset is the far-field scattering diagram of an electric dipole (the identical shape to each structure here).

**Outlook and Conclusion**

Electrically small antennas [38] and scatterers find use in many applications, including wireless communications. The main drawback of using small structures comes through their reduced scattering efficiencies and operational bandwidths compared to bigger counterparts. Quite a few limiting criteria have been derived over the years, with the Chu-Harrington limit being the most celebrated one. Although many proposals and demonstrations to surpass those limits have been reported, introducing additional degrees of freedom to electromagnetic designs can be beneficial. The field of metamaterials suggests employing tailored, effective material parameters to improve antennae and scatterers' performances.

Our current investigation aimed to test the metamaterial approach from the applied standpoint, which considers practical limitations. The surface equivalence principle allows replacing a volumetric scatterer with an impedance surface, which grants the same electromagnetic properties, including the far-field scattering. However, practical realizations, in many cases, can challenge the theory. Here, we investigated this question by comparing metamaterial-based scatterers with their surface counterparts. In particular, arrays of inductively loaded wires replicate materials with effective negative permittivity. Constructing subwavelength scatterers from the dipole-media allows emulating the phenomenon of localized plasmon resonance (e.g., [39], [40]) if $\varepsilon \approx -2$ condition for a deeply subwavelength sphere is satisfied. The surface equivalent of this structure is a helical spiral on a spherical surface. These two geometries (the spiral and the metamaterial) were compared in terms of scattering efficiencies and the bandwidth. We found that in the case of these simple geometries, the surface equivalent outperforms the metamaterial in both parameters. Metamaterials with finite-size unit cells suffer from additional losses associated with rough effective boundaries and internal material losses. The latter is associated with realistic lumped elements and substrates forming the structure.

On the other hand, the helical structure can be implemented with low-loss metals (copper) and has a very smooth form factor. While the surface equivalence principle predicts obtaining the same scattering peak as the volumetric structure, it does not make any prediction on the bandwidth. The bandwidth strongly depends on a specific realization, and it is inherently linked with geometric layout, material dispersion, or even both. It is worth noting that metamaterial designs seem to be more flexible in obtaining complex functionalities. In the context of this investigation, pushing the resonance of the helical structure to higher frequencies is rather challenging. At the same time, this task can be straightforwardly addressed if the metamaterial design is used.

Furthermore, in cases when higher-order spectrally overlapping resonances are required in the design, the advantages one of the approaches over another are still an open question. This type of analysis is highly required for the construction of super-directive antennas and super-scatterers. While the performed comparison is strongly linked to specific realizations (helical impedance surfaces and dipole-based metamaterial), the beforehand-discussed trends are quite general. Helical structures have been a topic for optimization for quite a while and demonstrate next to optimal performances, as demonstrated in, e.g., [28,35]. Effective permittivity in metamaterial realizations is typically based on dipole arrays. While unit cells can have quite diverse designs, the basic operation principle is the hybridization of dipolar modes.

In this case, the main limitation is the granularity associated with finite-size unit cells, which limits the performance, as demonstrated here.

**Acknowledgments**

The research was partly supported by the Pazy Foundation, ERC StG 'In Motion (802279),' the Ministry of Science and Technology (project "Integrated 2D&3D Functional Printing of Batteries with Metamaterials and Antennas").